# Geometric energy transfer in a Stückelberg interferometer of two parametrically coupled mechanical modes

Hao Fu,[1,3,#] Zhi-cheng Gong,[1,2,#] Tian-hua Mao,[1] Cheng-yu Shen,[1,2] Chang-pu Sun,[3,4] Su Yi,[5,6] Yong Li,[4, *] & Geng-yu Cao[1,*]

[1] *State Key Laboratory of Magnetic Resonance and Atomic and Molecular Physics, Wuhan Institute of Physics and Mathematics, Chinese Academy of Sciences, Wuhan 430071, China*

[2] *University of Chinese Academy of Sciences, Beijing 100190, China*

[3] *Graduate School of the China Academy of Engineering Physics, Beijing 100193, China*

[4] *Beijing Computational Science Research Center, Beijing 100193, China*

[5] *CAS Key Laboratory of Theoretical Physics, Institute of Theoretical Physics, Chinese Academy of Sciences, Beijing 100190, China*

[6] *School of Physics & CAS Center for Excellence in Topological Quantum Computation, University of Chinese Academy of Sciences, Beijing 100190, China*

[#] These authors contribute equally to this works

[*] Correspondence and requests for materials should be addressed to Y. L. (email: liyong@csrc.ac.cn) or G.-Y. C. (email: gycao@wipm.ac.cn)

**Abstract**

Geometric phase, which is acquired after a system undergoing cyclic evolution in the Hilbert space, is believed to be noise-resilient because it depends only on the global properties of the evolution path. Here, we report geometric control of energy transfer between two parametrically coupled mechanical modes in an optomechanical system. The parametric pump is controlled along a closed loop to implement geometric Stückelberg interferometry of mechanical motion, in which the dynamical phase is eliminated using the Hahn-echo technique. We demonstrate that the interference based solely on the geometric phase is robust against certain noises. And, more remarkably, we show that the all-geometric approach can achieve a high energy-transfer rate that is comparable to those using conventional dynamical protocols.

Strongly coupled resonators have been intensively investigated in various distinct physical systems [1-7], in which coherent control of the transport phenomena has been realized using the rich toolbox developed for the quantum systems. Specifically, for example, two-mode mechanical system has been demonstrated as a classical analogy of quantum two-level system, where the Rabi physics was applied for full control of mechanical motion in the Bloch sphere [7-10]. And more recently, studies in the two-mode mechanical system have shown that non-adiabatically traversing the anti-crossing point follows the Landau-Zener (LZ) dynamics, in which the coherent splitting [11,12], Stückelberg interference [12,13], and intriguing effect of dynamical localization [14] for the mechanical motion have been achieved experimentally. Besides the high potential on quantum information processing with respect to the recent great advances on cavity optomechanics [15-19], these dynamical protocols have been widely adopted in coherent control of energy transfer between mechanical resonators for applications such as coherent switches [14], low-power logic units [20,21], non-reciprocal transducers [22], and coherent force sensors [23]. A continuous challenge in developing the coherent mechanical devices and further extending their applications is improving the coherence of operations. Benefiting from the great advances on nano fabrications, mechanical resonators with extremely low dissipation have become available [24-26], which can significantly reduce energy dissipation during operations. Nevertheless, the coherent control of energy transfer in either optomechanical or electromechanical systems will be unavoidably subject to noises from various sources, which can cause observable phase decoherence [27,28].

So far, significant efforts have been devoted to developing noise-resilient protocols for coherent control of energy transfer in mechanical systems. For example, topological energy transfer between two modes of a mechanical resonator has been achieved recently through adiabatically encircling the exceptional point [29], where the effective Hamiltonian of the system is non-Hermitian. Alternatively, when a system undergoes cyclic evolution in the Hilbert space through adiabatically changing the control parameters, it acquires a geometric phase in addition to the dynamical one [30,31]. Because it depends only on the global properties of the evolution path, the geometric phase is believed to be robust against certain noises and has practical applications in fault-tolerant quantum information processing [32-35]. In order to overcome the operation speed limited by the adiabaticity, which typically requires the evolution in the parametric space at a rate significantly smaller than the typical coupling strength, numerous non-adiabatic geometric protocols have also been proposed [36-39]. And the quantum logic gates based on the non-adiabatic geometric phases have been extensively implemented in the systems of magnetic resonances [40], superconductor qubits [41,42], and trapped ions [43,44].

In this paper, we demonstrate energy transfer through geometric Stückelberg interferometry of two parametrically coupled mechanical modes. We show that the geometric phase can be acquired after the cyclic evolution of the motion state between the two consecutive LZ transitions. In order to realize an all-geometric operation, the dynamical phase accumulated during the cyclic evolution is eliminated using the Hahn-echo technique. We demonstrate that the noise-resilient energy transfer can be achieved using the geometric Stückelberg interferometry at the operation speed comparable with that in the conventional dynamical protocols.

The system under investigation is a tunable two-mode optomechanical system consisting of two cantilevers, which are elastically coupled by connecting to the same overhang. By trapping one of the cantilevers (cantilever 1) inside a fiber-based cavity using a 1,064 nm laser, the frequency of cantilever 1 becomes trapping power $P$ dependent $\omega_{1,\text{eff}} = \sqrt{\omega_1^2 + g\,P}$ with $g$ representing the strength of optical trap; while the frequency of cantilever 2, $\omega_2$, remains unaffected [45]. We use an additional weak 1,310 nm laser, which is linearly coupled to the mechanical motion, to monitor the oscillation of cantilever 1. Owing to the elastic coupling, the fundamental flexural modes of the cantilevers are hybridized into two normal modes with the two cantilevers oscillating in-phase ($x_-$) and out-of-phase ($x_+$). The strength of the elastic coupling measured as the frequency difference between two normal modes at degenerate point (where $\omega_{1,\text{eff}} = \omega_2$) is $\Delta/2\pi = 459$ Hz. Our experiments are carried out at the trapping power of $P_0 = 131\,\mu\text{W}$, at which the frequencies of the in-phase and out-of-phase modes are $\omega_-/2\pi = 6{,}234$ Hz and $\omega_+/2\pi = 6{,}701$ Hz, respectively (see supplemental material). Here, the system is intentionally offset from the degenerate point so that frequency fluctuation of the normal modes can be created through introducing extra noises on the trapping power $P$ in order to study the effect of noises on the operation as we will show later. As a result of the gentle cold-damping effect of the 1,310 nm probing laser, the dissipation rates of the two normal modes are slightly different with $\gamma_-/2\pi = 0.37$ Hz and $\gamma_+/2\pi = 0.25$ Hz.

In order to couple the two normal modes, a parametric pump is applied through modulating the trapping power $P(t) = P_0 + P_d\cos(\omega_d t + \theta)$. When the parametric pump is activated, the dynamics of periodically driven two-mode mechanical system can be described by

$$\begin{pmatrix} \frac{d^2}{dt^2} + \gamma_+\frac{d}{dt} + \omega_+^2 & 0 \\ 0 & \frac{d^2}{dt^2} + \gamma_-\frac{d}{dt} + \omega_-^2 \end{pmatrix}\begin{pmatrix} x_+ \\ x_- \end{pmatrix} - \varepsilon(t)\begin{pmatrix} 1 + \cos\alpha & \sin\alpha \\ \sin\alpha & 1 - \cos\alpha \end{pmatrix}\begin{pmatrix} x_+ \\ x_- \end{pmatrix} = 0, \quad (1)$$

where $\varepsilon(t) = \frac{1}{2}gP_d\cos(\omega_d t + \theta)$ denotes the parametric pump and $\alpha$ satisfies $\tan\alpha \approx \frac{\Delta}{\omega_{1,\text{eff}}(P_0) - \omega_2}$. When the frequency difference between the normal modes, $\delta\omega \equiv \omega_+ - \omega_-$, is compensated by the parametric pump, mixing of the two normal modes leads to a normal-mode splitting. The anti-crossing phenomenon can be observed in Fig. 1(a). And the strength of the parametric coupling between the two normal modes, $\Omega$, which is proportional to the pump power $P_d$, can be measured as the normal-mode splitting at the on-resonance condition $\omega_d = \delta\omega = 2\pi \times 467$ Hz [Fig. 1(b)]. In contrast to the previous Stückelberg interferometer based on two mechanical modes coupled through interacting with the same static field [12], we use the two parametrically-coupled normal modes to construct a Stückelberg interferometer, in which the coupling field is fully tunable.

To overcome the thermal Brownian motion, the system is initialized by piezo-electrically actuating the in-phase mode. When the anti-crossing point is traversed non-adiabatically, it functions as a coherent splitter to transfer part of energy from the in-phase mode to the out-of-phase mode. As the first step of the Stückelberg interferometry, we implement the coherent splitter through applying a frequency-modulated pump pulse with the pump frequency ($\omega_d$) ramped linearly from $\omega_a = 2\pi \times 267$ Hz to $\omega_b = 2\pi \times 667$ Hz in the transition time $t_{\text{LZ}}$. The energy splitting ratio, which is measured as the proportion of energy remaining on the in-phase mode after the transition, can be calculated using the LZ formula [46,47]

$$P_{\text{LZ}} = \exp\left(-\frac{\pi\Omega^2}{2\nu}\right), \quad (2)$$

with $\nu = \frac{|\omega_a - \omega_b|}{t_{LZ}}$ denoting the speed at which the anti-crossing point is traversed. For the parametric coupling strength $\Omega/2\pi = 28.5$ Hz, the oscillation amplitudes of the in-phase ($X_-$) and out-of-phase ($X_+$) modes are recorded simultaneously after the pump pulse (see supplemental material). As shown in Fig. 1(c), the energy splitting ratio can be tuned according to equation (2) through changing the transition time $t_{LZ}$. And, specifically, a 50/50 splitter is achieved with the transition time $t_{LZ} \approx 30$ ms.

For a round-trip LZ transition, the Stückelberg interferometry of two parametrically coupled mechanical modes can be achieved. The recombination of the mechanical motions after the second LZ transition creates an interference fringe depending on the relative phase of the motions. Generally, the phase acquired between the two consecutive LZ transitions includes both the dynamical phase and the geometric phase that depends solely on the evolution path of the system in state space [48]. In order to realize the Stückelberg interferometry based solely on the geometric phase, the dynamical phase is eliminated in our experiment using the Hahn-echo technique. We implement the all-geometric energy transfer through modulating the parametric pump $\varepsilon(t) \to \frac{1}{2}gP_d(t)\cos[\int \omega_d(t')dt' + \theta(t)]$, where the power $P_d(t)$, frequency $\omega_d(t)$, and phase $\theta(t)$ of the pump are controlled as schematically illustrated in Fig. 2(a). Our protocol is similar with that in Ref. [42] except that the pump detuning $\Omega_z \equiv \delta\omega - \omega_d$ is swept via modulating the pump frequency $\omega_d$ in this work rather than changing $\delta\omega$ as done in Ref. [42]. Although our system is intentionally offset from the degenerate point ($\delta\omega \neq \Delta$), we note that this modification is non-trivial because it allows applying the geometric Stückelberg interferometry even at the degenerate point, where $\delta\omega$ is most stable against the fluctuation in control parameter, namely $\frac{\partial \delta\omega}{\partial P} \sim 0$ in our case.

The evolution of the two-mode mechanical system can be described on the Bloch sphere with the longitude field $\Omega_z$ and transverse fields $\Omega_x = \Omega\cos(\theta)$ and $\Omega_y = \Omega\sin(\theta)$, respectively. After the initial state of system is prepared to the in-phase mode, the pump pulse is activated ($t = t_A = 0$). The loop path of the parametric pump is shown in Fig. 2(b). According to the adiabatic-impulse approximation [48], the system evolves adiabatically except at the time instant when the anti-crossing point is traversed, where the mechanical motion is split as a consequence of the LZ transition. Therefore, after the first LZ transition at $t = t_C$, the motion of the system can be described as a superposition of two normal modes, which are marked as two orthogonal motion states C and C' on the Bloch sphere [see Fig. 2(c)]. We note that the adiabatic evolution of the system between two LZ transitions can be described by the pair of orthogonal cyclic states C and C'. In what follows, we just focus on the evolution path of the state C, which is plotted on the Bloch sphere in Fig. 2(c). The trajectory CE of evolution adiabatically follows the pump field trajectory CDE in Fig. 2(b). At the time instant $t_E$, a π pulse is applied to flip the Bloch vector from the point E to the point F around the *x* axis. Then, the phase of the pump is reversed to bring the system back to the point C through the trajectory FH on the Bloch sphere. The durations for the evolution CE and FH are kept to be equal ($t_{FH} = t_{CE}$) so that the dynamical phase, including both the Stokes phase acquired at each LZ transition and the adiabatic dynamical phase accumulated during the evolution CE and FH, is completely cancelled out after a full evolution cycle. Consequently, the recombination of the mechanical motions as a result of the second LZ transition at $t = t_H$ creates an interference fringe depending solely on the geometric phase $\phi$, which can be calculated as half of the solid angle enclosed by the trajectory CEFH on the Bloch sphere. And the oscillation amplitude of the out-of-phase mode after the round-trip transition can be calculated as

$$X_+ = \sqrt{1 - 4P_{\text{LZ}}(1 - P_{\text{LZ}})\cos^2\phi}. \tag{3}$$

The geometric Stückelberg interferometry is highly controllable since the energy splitting ratio $P_{\text{LZ}}$ and the geometric phase $\phi$ can be adjusted independently through controlling the pump field. As shown in Fig. 2(d), an all-geometric energy transfer is achieved through changing the phase $\phi$. To enhance the transfer efficiency, the energy splitting ratio is optimized for a 50/50 splitter through adjusting the transition time. In particular, for the parametric coupling strength $\Omega(t_{C,H})/2\pi = 28.5$ Hz, performing the geometric Stückelberg interferometry at the transition time $t_{BD} = t_{GI} = 30$ ms yields an interference with its visibility reaching approximately 0.94 [49]. For comparison, a dynamical interference is implemented using the protocol similar with that illustrated in Fig. 2(a) except that the π pulse is removed in order to preserve the dynamical phase. We control the dynamical phase through adjusting the evolution time $t_{DE} = t_{FG}$ while keeping the geometric phase $\phi = 0$ as constant. Generally, a precise control of the dynamical interference can be expected providing the frequencies of normal modes are measured accurately. Otherwise, the error in controlling the dynamical phase can be accumulated with the evolution time increasing and lead to the observed mismatch between the experimental and the theoretical results in Fig. 2(e). Here, each data point in Figs. 2 (d) and (e) is the average result of seven independent measurements using the same parameters. Therefore, for different measurements, the fluctuation of the phase acquired between the two LZ transitions can be reflected as the deviation of these measurements in the figures. It is interesting to see that the effect of decoherence originating from the dephasing is more significant for the case of dynamical interference compared with that of the geometric one.

The improved coherence obtained in the geometric Stückelberg interferometry implies that our protocol is more robust against certain noises than the conventional dynamical one. To investigate the effect of noise on the operation coherence, two kinds of white noises of the same amplitude are intentionally added in the trapping power to create random fluctuations on the pump power with the bandwidth of 10 Hz and 1000 Hz, respectively. For the noise with bandwidth of 10 Hz, although the pump field can be treated approximately as uniform during each measurement, it causes a slow fluctuation of pump field compared with the adiabatic evolution time scale $t_{CH}$ so that the pump field is inhomogeneous for different measurements. By increasing the noise bandwidth to 1000 Hz, we also include noise at high frequency to induce fast fluctuation of pump field during each measurement. As shown in Fig. 3(a), both the noises can increase decoherence in the dynamical interference. And, moreover, the noise-induced phase fluctuation can be accumulated and the effect of dephasing is amplified with the increase of operation time. From the interference fringes, we can obtain the standard deviation of phase, $\delta\phi$, accumulated between the two consecutive LZ transitions. The average values of $\delta\phi$ for the adiabatic evolution time between $t_{CH} = 45$ ms and $t_{CH} = 51$ ms are calculated to be 0.83 rad and 0.86 rad for noises with bandwidth of 10 Hz and 1000 Hz, respectively. In contrast, a noise-resilient energy transfer is achieved using the geometric Stückelberg interferometry [Fig. 3(b)]. For the measurements with noise bandwidth of 10 Hz, we note that the decoherence induced by the slow fluctuation of pump field is significantly reduced by the Hahn-echo technique, which is used to cancel the dynamical phase accumulated between the two consecutive LZ transitions. And, more remarkably, it is interesting to see that the well-preserved coherence can also been obtained under the influence of the noise with bandwidth of 1000 Hz. Although the low-frequency noise in this bandwidth can be effectively suppressed using the Hahn echo, the resilience to the noise that also includes high frequency components should not be completely attributed to the use of Hahn echo. The well-preserved coherence obtained even in present of the fast fluctuating pump field offers clear evidence that the geometric phase is robust against local fluctuation on the evolution path. The phase deviations as calculated from the geometric interference fringe are nearly equal for the two noises, with its average value $\delta\phi \approx 0.10$ rad.

In summary, we have presented the geometric control of energy transfer in the Stückelberg interferometer consisting of two parametrically coupled mechanical modes. When the parametric pump is on-resonance, an anti-crossing phenomenon is observed which allows for energy transfer between the mechanical modes. Our experiments demonstrate that the non-adiabatic transition of the anti-crossing point through modulating the pump frequency leads to coherent splitting of mechanical motion following the LZ model. For a round-trip LZ transition, the all-geometric control of energy transfer based on the Stückelberg interferometry has been implemented through eliminating the dynamical phase using the Hahn-echo technique. We note that the speed of geometric operation is comparable with the typical coupling strength ($2\pi/t_{AJ} \sim \Omega$). Moreover, we show that our protocol integrating the merits of both the geometric operation, which is inherently robust against local fluctuations on the control parameters, and the Hahn-echo technique, which is capable of cancelling the slow fluctuations, can achieve high efficient transfer of mechanical energy with a well-preserved phase coherence even in the presence of certain noises. Therefore, we conclude that our protocol offers a fast and highly controllable method for noise-resilient control of mechanical energy transfer, which can be generally implemented in most electro-mechanical and opto-mechanical systems using currently available technology.

**Acknowledgment**

This work was supported by MOST under Grant No. 2017YFA0304500, Science Challenge Project under Grant No. TZ2018003, and NSFC under Grants No. 91636220, No. 11774024, No. 11534002, and No. 11434011. We would like to thank nanofabrication facility in Suzhou Institute of Nanotech and Nano bionics (CAS) for fabricating the micro-cantilevers.

**Figure 1:**

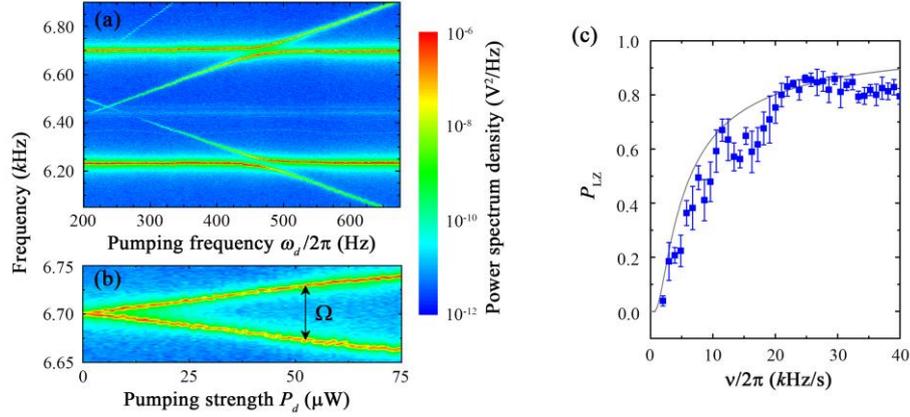

**Figure 1 Parametrically coupled two mechanical modes. (a)** The pumping frequency response of the parametrically coupled mechanical modes. The parametric pump is applied by modulating the trapping power at the amplitude $P_d = 29~\mu W$. The thermal oscillation power spectral density of cantilever 1 under the parametric pump is measured using an electro-spectra analyzer. **(b)** The modulating amplitude $P_d$ dependent normal-mode splitting at the on-resonance condition. For each modulating amplitude, the strength of parametrical coupling $\Omega$ is measured as the anti-crossing gap as marked in panel **(b)**. **(c)** Energy splitting ratio for each transition speed. Here, the energy splitting ratio $P_{\text{LZ}}$ measured as the relative oscillation energy of the in-phase mode after the pump pulse (solid blue dots), $\frac{X_-^2}{X_-^2 + X_+^2}$, is plotted with the theoretical result (gray line) calculated from equation (2).

**Figure 2:**

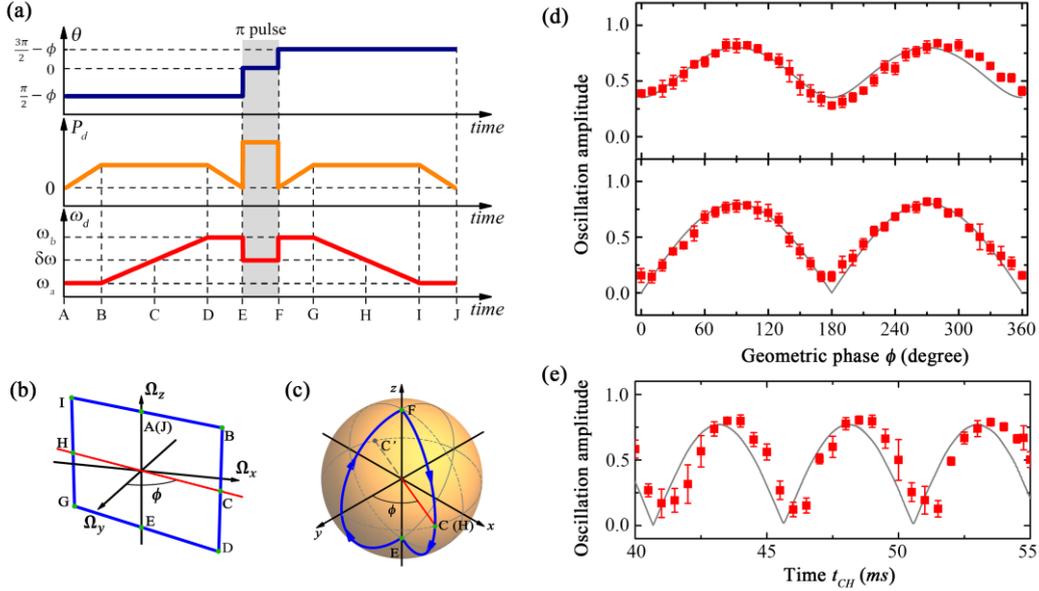

**Figure 2 Stückelberg interferometry with two parametrically coupled modes. (a)** The control sequence of the pump pulse for the geometric Stückelberg interferometry. The strength of parametric coupling is ramped up to $\Omega/2\pi = 28.5$ Hz in $t_{AB} = 5$ ms, and after the pump frequency swept from $\omega_a/2\pi = 267$ Hz to $\omega_b/2\pi = 667$ Hz for the first LZ transition the parametric pump is gradually turned off in $t_{DE} = 5$ ms. After the $\pi$ pulse, the parametric pump with its phase shifted by 180 degree is applied for the second LZ transition at the same speed ($t_{BD} = t_{GI}$). The parametric pump is gradually turned on (off) before (after) the transition in $t_{FG} = t_{IJ} = 5$ ms. And the width of the $\pi$ pulse is $t_{EF} \approx 8.6$ ms. **(b)** The trajectory of the effective parametric field in parametric space. The points marked on the loop path correspond to the parametric pump at the time instants marked in panel **(a)**. **(c)** The evolution path of the system on the Bloch sphere. The points marked on the Bloch sphere correspond to the motion states of in-phase mode at time instants marked in panel **(a)**. **(d)** Geometric Stückelberg interference. The final oscillation amplitude of the out-of-phase mode is measured for the transition times $t_{BD} = t_{GI} = 15$ ms (upper panel) and $t_{BD} = t_{GI} = 30$ ms (lower panel), respectively. For each geometric phase $\phi$, the oscillation amplitudes of the out-of-phase mode (solid red dots) are plotted with the theoretical results (gray lines) calculated from equation (3). **(e)** Dynamical Stückelberg interference. For the transition time $t_{BD} = t_{GI} = 30$ ms, the oscillation amplitudes of the out-of-phase mode measured after the pump pulse (solid red dots) and calculated from equation (1) (gray lines) are plotted against the total evolution time between the two LZ transitions $t_{CH}$.

**Figure 3:**

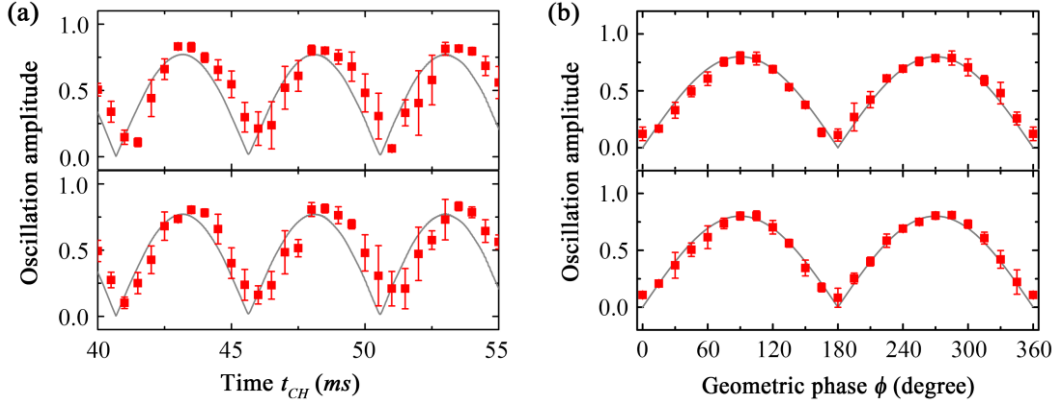

**Figure 3 Effect of the noises on the Stückelberg interference.** (**a**) Dynamical Stückelberg interference and (**b**) geometric Stückelberg interference in the presence of the random noises with bandwidth 1000 Hz (upper panel) and 10 Hz (lower panel). The white noise is added in the pump power to create random fluctuation on the strength of longitude field with the amplitude approximately 3 Hz in peak. The similar pump pulses used in Figs. 2(d) and (e) are applied respectively to realize the geometric and dynamical Stückelberg interferometry with the parametric coupling strength at each LZ transition $\Omega(t_{C,H})/2\pi = 28.5$ Hz and the transition time $t_{BD} = t_{GI} = 30$ ms. The oscillation amplitudes of the out-of-phase mode measured after the operation (solid red dots) are plotted. The corresponding numerical results are also given (gray lines) according to equation (1).

# Supplemental Material for "Geometric energy transfer in a Stückelberg interferometer of two parametrically coupled mechanical modes"

## A. Experimental setup

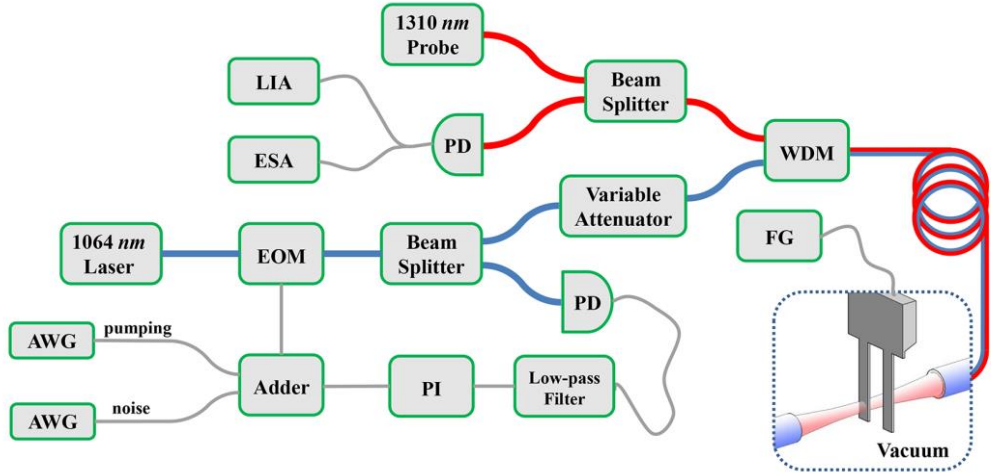

**Figure S1 Schematic illustration of experimental setup.** The motion of the cantilever is analyzed by a lock-in amplifier (LIA) and electro spectrum analyzer (ESA) through measuring the intensity of the reflected 1,310 nm laser. An additional 1,064 nm laser, which is combined with the 1,310 nm laser after the wavelength division multiplexer (WDM), is used as a trapping laser. The power of the 1,064 nm laser is monitored by a photo detector (PD) and stabilized by a PI-type feedback loop. During the parametric pump is activated, the feedback is hold. A function generator (FG) is used to drive a piezo-electric actuator (not shown on the figure) attached on the substrate of the cantilevers in order to initialize the system.

The mechanical resonators used in our experiments are two single-crystal silicon cantilevers with dimension of 220 μm in length, 10 μm in width, and 200 nm in thickness. The silica substrate is etched so that the two cantilevers separated by 15 μm are elastically coupled by connecting to the same thin silicon overhang extended 15 μm outward from the substrate. In order to mediate the effective coupling between the cantilevers, one of the cantilevers (cantilever 1) is inserted into a low finesse cavity, which is formed by two 15 nm gold coated plane fibers. The laser beam from the fiber is focused to about 3 μm in radius using a micro-lens. And three electric-piezo stages are used to align the cavity and the cantilever. As illustrated in Fig. S1, the cavity is pumped by a weak 1,310 nm laser and a 1,064 nm laser with its power $P$ tunable via an electro-optic modulator (EOM). We carefully adjust the cavity length so that the 1,310 nm and 1,064 nm lasers are coupled linearly and quadratically to the mechanical motion, respectively. Owing to the quadratic optomechanical coupling [1], the 1,064 nm laser introduces an additional optical spring to the cantilever 1, $k_{opt} = g\,P$, with $g$ representing the optical trapping strength. Therefore, the frequency of cantilever 1 becomes trapping power dependent, $\omega_{1,\text{eff}}(P) = \sqrt{\omega_1^2 + g\,P}$, while the frequency of the other cantilever (cantilever 2) $\omega_2$ remains unaffected.

To overcome the thermal oscillation, the in-phase mode is actuated piezo-electrically so that it oscillates at amplitude significantly larger than that of its thermal Brownian motion. Immediately after the initialization ($t = 0$), a parametric pump pulse as described in the main text is applied. In our experiment, an arbitrary waveform generator (AWG) is used to synthesize the driving signal for the EOM to create the parametric pump through modulating the intensity of trapping laser. And a white noise generated by another AWG can be added on the pump power to intentionally create random fluctuation of the pump field.

**B. Non-adiabatic dynamics of the parametrically coupled two mechanical modes**

The motion of the cantilevers under consideration, $x_1$ and $x_2$, can be described as

$$\begin{pmatrix} \frac{d^2}{dt^2} + \gamma_1 \frac{d}{dt} + \omega_{1,eff}^2(P) & J \\ J & \frac{d^2}{dt^2} + \gamma_2 \frac{d}{dt} + \omega_2^2 \end{pmatrix} \begin{pmatrix} x_1 \\ x_2 \end{pmatrix} = 0, \tag{S1}$$

where $\gamma_i$ is the dissipation rates of the $i^{th}$ cantilever, and $J$ represents the strength of elastic interaction between the cantilevers. Owing to the elastic coupling, the motion of the cantilevers is hybridized into two normal modes. In the case of $\gamma_i \ll \omega_i$, we can diagonalize Eq. (S1) by neglecting the mechanical damping effect and obtain the motion of in-phase ($x_-$) and out-of-phase ($x_+$) modes $\begin{pmatrix} x_+ \\ x_- \end{pmatrix} = U \begin{pmatrix} x_1 \\ x_2 \end{pmatrix}$ with its eigenfrequency

$$\omega_\pm^2(P) = \frac{1}{2}\left(\omega_{1,eff}^2(P) + \omega_2^2 \pm \sqrt{\left(\omega_{1,\text{eff}}^2(P) - \omega_2^2\right)^2 + 4J^2}\right), \tag{S2}$$

where $U(P) = \begin{pmatrix} u_{11} & u_{12} \\ u_{21} & u_{22} \end{pmatrix} = \begin{pmatrix} \cos\frac{\alpha}{2} & \sin\frac{\alpha}{2} \\ -\sin\frac{\alpha}{2} & \cos\frac{\alpha}{2} \end{pmatrix}$ and $\alpha$ satisfying $\tan\alpha = \frac{2J}{\omega_{1,\text{eff}}^2(P) - \omega_2^2}$. The strength of the elastic coupling, which is measured as the frequency splitting of the normal modes at the degenerate point ($\omega_{1,eff} = \omega_2$), can be calculated approximately according to $\Delta \approx J/\omega_2$ in the case of $J \ll \omega_2^2$. We mediate the hybridization of the cantilevers through tuning the trapping power $P$. As shown in Fig. S2, when the trapping power is swept adiabatically, an avoided crossing of the two normal modes can be clearly observed. The minimal frequency splitting between the normal modes $\Delta/2\pi = 459$ Hz can be obtained at the degenerate point with the trapping power reaching 182 μW. Fitting the measured resonant frequencies of the normal modes to Eq. (S2), we obtain the optical trapping strength $g \approx -9.8 \times 10^5\ rad^2/(s^2 \cdot \mu W)$. We carry out the experiments at the trapping power $P_0 = 131\ \mu W$, at which the resonant frequencies of the in-phase and out-of-phase modes are $\omega_-/2\pi = 6{,}234$ Hz and $\omega_+/2\pi = 6{,}701$ Hz, respectively.

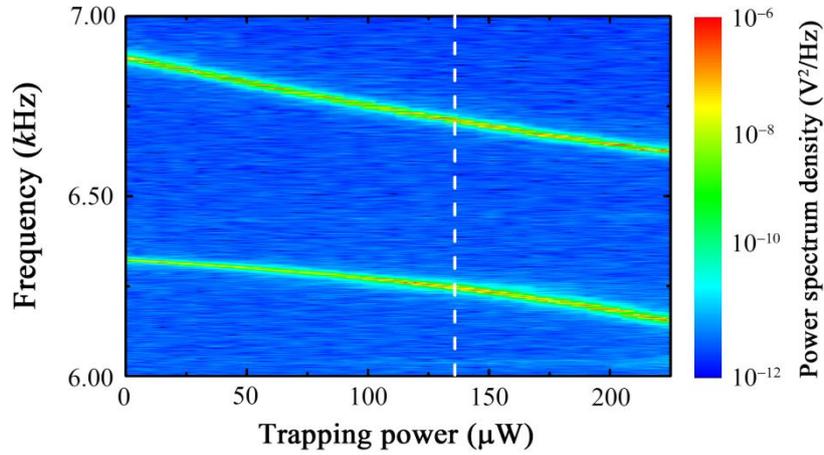

**Figure S2 Thermal oscillation power spectrum density of cantilever 1.** The trapping power at which the experiments are carried out is marked by the dashed line.

When the trapping power is periodically modulated as $P(t) = P_0 + P_d\cos(\omega_d t + \theta)$, the two normal modes become parametrically coupled with their motion satisfying

$$\begin{pmatrix} \frac{d^2}{dt^2} + \gamma_+ \frac{d}{dt} + \omega_+^2 & 0 \\ 0 & \frac{d^2}{dt^2} + \gamma_- \frac{d}{dt} + \omega_-^2 \end{pmatrix} \begin{pmatrix} x_+ \\ x_- \end{pmatrix} + \varepsilon(t) \begin{pmatrix} 1+\cos\alpha & -\sin\alpha \\ -\sin\alpha & 1-\cos\alpha \end{pmatrix} \begin{pmatrix} x_+ \\ x_- \end{pmatrix} = 0, \quad (S3)$$

where $\gamma_\pm$ is the energy dissipation rate of the normal mode and $\varepsilon(t) = \frac{1}{2} g P_d \cos(\omega_d t + \theta)$ denotes the parametric pump. Generally, the motion of the normal modes in the presence of the parametric pump can be written as

$$x_+(t) = Re[X_+(t) e^{-i\omega_+ t}],$$

$$x_-(t) = Re[X_-(t) e^{-i\omega_- t}], \quad (S4)$$

where $X_\pm(t)$ is the complex oscillation amplitude. Generally, $X_\pm(t)$ contains all idle components that oscillate at frequencies of integral multiplex of $\omega_d$ [2]. Here, we focus only on the slowly varying components that contribute to the inter-mode energy transfer as we will show later, and assume $X_\pm(t)$ varies slowly as comparing with the resonant frequency $\omega_\pm$ of the modes. For the high-quality mechanical resonators ($\gamma_\pm \ll \omega_\pm$), we can substitute Eq. (S4) into Eq. (S3) and obtain

$$\begin{pmatrix} -2i\omega_+ \frac{d}{dt} + 2\varepsilon(t)\cos^2\frac{\alpha}{2} & -\varepsilon(t)\sin\alpha \, e^{i\delta\omega t} \\ -\varepsilon(t)\sin\alpha \, e^{-i\delta\omega t} & -2i\omega_- \frac{d}{dt} + 2\varepsilon(t)\sin^2\frac{\alpha}{2} \end{pmatrix} \begin{pmatrix} X_+(t) \\ X_-(t) \end{pmatrix} = 0, \quad (S5)$$

where $\delta\omega = \omega_+ - \omega_-$. Performing the transform $X_+(t) = a_+(t) e^{-\frac{i}{\omega_+}\cos^2\frac{\alpha}{2}\int \varepsilon(t')dt'}$ and $X_-(t) = a_-(t) e^{-\frac{i}{\omega_-}\sin^2\frac{\alpha}{2}\int \varepsilon(t')dt'}$ and using expansion $e^{\pm i\Lambda \sin(\omega_d t+\theta)} = \sum_q (-1)^q J_q(\Lambda) e^{\mp iq(\omega_d t+\theta)}$, Eq. (S5) can be written as

$$i\sqrt{\frac{\omega_+}{\omega_-}} \frac{da_+(t)}{dt} + \frac{\Omega}{2}\left[e^{i(\omega_d t+\theta)} + e^{-i(\omega_d t+\theta)}\right] e^{i\delta\omega t} \sum_q (-1)^q J_q(\Lambda) e^{-iq(\omega_d t+\theta)} a_-(t) = 0,$$

$$i\sqrt{\frac{\omega_-}{\omega_+}} \frac{da_-(t)}{dt} + \frac{\Omega}{2}\left[e^{i(\omega_d t+\theta)} + e^{-i(\omega_d t+\theta)}\right] e^{-i\delta\omega t} \sum_q (-1)^q J_q(\Lambda) e^{iq(\omega_d t+\theta)} a_+(t) = 0, \quad (S6)$$

where $\Omega = \frac{g P_d \sin\alpha}{4\sqrt{\omega_-\omega_+}}$ represents the strength of parametric coupling and $\Lambda = \frac{g P_d \left(\omega_- \cos^2\frac{\alpha}{2} - \omega_+ \sin^2\frac{\alpha}{2}\right)}{2\omega_-\omega_+\omega_d}$.

The system is on-resonance when the pump frequency satisfies $\omega_d = \delta\omega/n$ with $n=1,2,3,\ldots$. Therefore, at the on-resonance condition, Eq. (S6) can be reduced by neglecting the fast oscillating off-resonance terms

$$i\sqrt{\frac{\omega_+}{\omega_-}}\frac{da_+(t)}{dt} + \frac{(-1)^{n-1}}{2}\Omega[J_{n-1}(\Lambda) + J_{n+1}(\Lambda)]e^{-in\theta}a_-(t) = 0,$$

$$i\sqrt{\frac{\omega_-}{\omega_+}}\frac{da_-(t)}{dt} + \frac{(-1)^{n-1}}{2}\Omega[J_{n-1}(\Lambda) + J_{n+1}(\Lambda)]e^{in\theta}a_+(t) = 0, \quad (S7)$$

where $J_n(\Lambda)$ is the $n$-th order Bessel function of the first kind. Providing the pump is weak ($|\Lambda| \ll 1$), which is the case in our experiments where $\Lambda \sim 0.02$, the contribution of higher-order component to the coherent energy transfer is negligible in Eq. (S7) since $J_{n-1}(\Lambda) \gg J_{n+1}(\Lambda)$. When the pump frequency is swept, we note that the dynamics of the two-mode mechanical system might involve processes with different $n$. To avoid the dynamics involving higher-order processes with $n > 0$, the Landau-Zener transition is implemented by sweeping the pump frequency from $\omega_a/2\pi = 267$ Hz to $\omega_b/2\pi = 667$ Hz. Therefore, only the process with $n = 1$ is considered in our experiments. And Eq. (S6) can be further reduced

$$\begin{pmatrix} i\sqrt{\frac{\omega_+}{\omega_-}}\frac{d}{dt} & \frac{\Omega}{2}e^{i(\Omega_z t - \theta)} \\ \frac{\Omega}{2}e^{-i(\Omega_z t - \theta)} & i\sqrt{\frac{\omega_-}{\omega_+}}\frac{d}{dt} \end{pmatrix}\begin{pmatrix} a_+(t) \\ a_-(t) \end{pmatrix} = 0, \quad (S8)$$

with the pump detuning $\Omega_z = \delta\omega - \omega_d$. At the on-resonance condition $\Omega_z = 0$, the parametric coupling strength $\Omega$ can be measured as the normal-mode splitting at the anti-crossing point as shown in Fig. 1(b).

We show that non-adiabatic traversing the anti-crossing point in the two-mode mechanical system is analogous to the LZ transition in quantum two-level system [3]. The non-adiabatic dynamics of the parametrically coupled mechanical modes are analyzed by numerically solving Eq. (S1) and Eq. (S8) using the parameters exactly the same as that in our experiments. In Fig. S3, the well agreement of the results from numerical calculation and the results calculated from Eq. (2) in the main context reveals that the non-adiabatic dynamics of the parametrically coupled mechanical modes follows the LZ model.

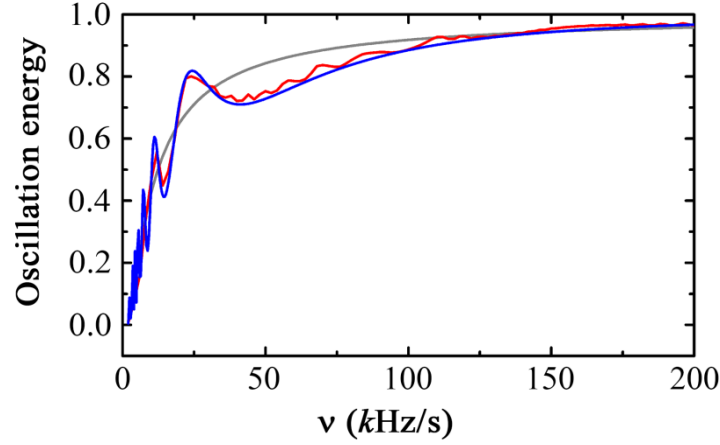

**Figure S3 Oscillation energy of in-phase mode.** The initial state of the system is prepared on the in-phase mode. And the final oscillation energy of the in-phase mode after the LZ transition is normalized to the total oscillation energy on the two normal modes. The final oscillation energy of the in-phase mode calculated from Eq. (S1) (red line) and Eq. (S8) (blue line) are plotted with the result calculated from Eq. (2) (gray line). The parameters used in the calculations are identical to that in our experiments.

**C. Data acquisition**

The experiments are performed in vacuum condition at room temperature. The oscillation of each normal mode is monitored through measuring the motion signal of cantilever 1. Because the motion of each cantilever is a superposition of two normal modes $\begin{pmatrix} x_1 \\ x_2 \end{pmatrix} = U \begin{pmatrix} x_+ \\ x_- \end{pmatrix}$, the two normal modes can be monitored by measuring the motion of cantilever 1. And two resonances corresponding to the normal modes can be discriminated in frequency domain on the oscillation spectrum of cantilever 1 (Fig. S2). At the trapping power of $P_0 = 131\mu W$, the relative contribution of the two normal modes to the motion of cantilever 1, $U(P_0) = \begin{pmatrix} u_{11} & u_{12} \\ u_{21} & u_{22} \end{pmatrix} = \begin{pmatrix} 0.78 & 0.62 \\ -0.62 & 0.78 \end{pmatrix}$, can be calculated. In our experiments, the oscillation amplitudes of the two normal modes are monitored simultaneously by measuring the amplitude $V_\pm(t)$ of the motion signal at the frequency $\omega_\pm$ using two lock-in amplifiers. The final oscillation amplitudes of the in-phase mode ($X_- = \frac{V_-(t_m)}{V_-(0)}$) and out-of-phase mode ($X_+ = \frac{|u_{21}|V_+(t_m)}{|u_{11}|V_-(0)}$) are obtained through normalizing the signal amplitude $V_\pm(t_m)$ measured at $t_m = 180$ ms to the amplitude $V_-(0)$ recorded immediately after the initialization by taking the relative contribution of each normal mode into account. The energy splitting ratio after a single-passage LZ transition is calculated as the proportion of oscillation energy on the in-phase mode $\frac{X_-^2}{X_-^2 + X_+^2}$ [4,5]. For a round-trip LZ transition, recombination of the oscillation at the second LZ transition leads to the Stückelberg interference. The visibility of interference in Fig. 2(d) is calculated as $\frac{X_{+,max}^2 - X_{+,min}^2}{X_{+,max}^2 + X_{+,min}^2}$, where $X_{+,max}$ and $X_{+,min}$ are the maximum and minimum amplitudes of the out-of-phase mode measured after the interference respectively [6].